\documentclass[epj]{svjour}

\usepackage{graphicx}
\usepackage{fancyhdr}
\def\makeheadbox{{
 }}
\usepackage{amsmath,amssymb,amsfonts}

\newcommand{\CiteSeeSaw}{\cite{Minkowski:1977sc,Yanagida:1980,Glashow:1979vf,Gell-Mann:1980vs,Mohapatra:1980ia}}

\newcommand{\mD}{\ensuremath{m_\text{D}}}
\newcommand{\mR}{\ensuremath{m_\text{R}}}
\newcommand{\nuR}{\nu_\text{R}}

\newcommand{\eV}{\ensuremath{\:\text{eV}}}
\newcommand{\GeV}{\ensuremath{\:\text{GeV}}}

\newcommand{\Eqref}[1]{Eq.~\eqref{#1}}

\def\mytitle{My title} 
\def\myauthors{My name}  
\def\mytype{My type of session}
\def\mysession{My session}

\def\mytitle{Can LHC Test the See-Saw Mechanism?} 
\def\myauthors{J\"orn Kersten}    
\def\mytype{Contributed Talk}    
\def\mysession{Flavor Physics}

\begin{document}
\title{Can LHC Test the See-Saw Mechanism?%
\thanks{Talk presented at the $15^\text{th}$ International Conference
 on Supersymmetry and the Unification of Fundamental Interactions
 (SUSY07), July 26 -- August 1, 2007, Karlsruhe, Germany.
 Based on work done in collaboration with Alexei Smirnov
 \cite{Kersten:2007vk}.
}
}
\author{J\"orn Kersten%
\thanks{\emph{Email:} jkersten@ictp.it}%
}                     
\institute{The Abdus Salam ICTP, Strada Costiera 11, 34014 Trieste, Italy
}
%
\date{}
\abstract{
We discuss the prospects for detecting right-handed
neutrinos which are introduced in the see-saw mechanism at future
colliders.  This requires
a very accurate cancellation between
contributions from different right-handed neutrinos to the light
neutrino mass matrix.  We search for
possible symmetries behind this cancellation and find that they have to
include lepton number conservation.  Light neutrino masses can
be generated as a result of small symmetry-breaking perturbations.
The impact of these perturbations on LHC physics is
negligible, so that the mechanism of neutrino mass generation and
LHC physics are decoupled in general.  In constrained cases,
accelerator observables and neutrino masses and mixings can be
correlated.
} 
\maketitle
\section{Introduction}
\label{sec:intro}

The (type-I) see-saw mechanism \CiteSeeSaw{} generates small neutrino
masses in a natural way, introducing
right-handed (RH) neutrinos that are singlets under the
Standard Model (SM) gauge group and can therefore have large Majorana
masses.  The light neutrino mass matrix is approximately given by
\begin{equation} \label{eq:SeeSaw}
	m_\nu = -\mD \mR^{-1} \mD^T \;,
\end{equation}
where $\mD$ is the Dirac mass matrix and $\mR$ is the Majorana mass
matrix of the heavy singlets.  A direct test of the see-saw mechanism
requires the detection of these heavy neutrinos and the measurement of
their Yukawa couplings.  Using \Eqref{eq:SeeSaw} in the case of only one
generation and $m_\nu \sim 0.1\eV$, we obtain the estimate $\mR \sim 10^{14}\GeV$,
if the Dirac neutrino masses are close to the electroweak scale.
The singlets may have masses as small as $100\GeV$, within the energy
reach of the LHC and other future colliders, if the Dirac masses are a
bit smaller than the electron mass, which does not appear completely
unreasonable either.  However, the RH
neutrinos interact with the SM particles only via Yukawa couplings,%
\footnote{This is the case in the minimal extension of the SM we
 consider here.  Of course, the situation is very different if the RH
 neutrinos have additional interactions, for example with TeV-scale
 SU(2)$_\text{R}$ gauge bosons.
}
which are tiny in this case.  Thus, we expect the RH neutrinos to be
either way too heavy or way too weakly coupled to be observable at
colliders.

However, this conclusion can be avoided if there are two or more RH
neutrinos
\cite{Wyler:1982dd,Bernabeu:1987gr,Buchmuller:1990du,Buchmuller:1991tu,Datta:1991mf,Pilaftsis:1991ug,Ingelman:1993ve,Heusch:1993qu,Tommasini:1995ii,Gluza:2002vs,Pilaftsis:2004xx,Pilaftsis:2005rv,Akhmedov:2006de,deGouvea:2007uz}.
Their contributions to the light neutrino masses can cancel, opening up the
possibility of rather light singlets with large Yukawa couplings but
exactly vanishing light neutrino masses.  Non-vanishing masses are
generated by small perturbations of the cancellation structure.
In this setup, the RH neutrinos may be observable in future collider
experiments.  This possibility has attracted renewed interest recently,
see e.g.\
\cite{deGouvea:2007uz,Han:2006ip,delAguila:2006dx,Atwood:2007fr,Bray:2007ru,deAlmeidaJr.:2007gc,delAguila:2007em}.

In the following, we will discuss the prospects for discovering RH
neutrinos at colliders from the point of view of theory.
We will consider the cancellation of contributions to the light
neutrino mass matrix and possible underlying symmetries in the next section.
After briefly discussing small perturbations of the
lead\-ing-order mass matrices that yield viable masses for the light
neutrinos, we will turn to consequences for signatures at colliders.
Within the setups relying on a symmetry, lepton number violation is
unobservable, while lepton-flavour-violating processes can have sizable
amplitudes.  Finally, we will comment on the implications a
detection of RH neutrinos would have for our understanding of the mechanism of neutrino mass generation.

\section{Cancellations and Symmetries}
\label{sec:CancelSym}

\subsection{Vanishing Light Masses}
\label{sec:Cancel} 
For three generations of left- and right-handed neutrinos, the
contributions of the RH neutrinos to the light mass matrix cancel
exactly, if and only if
\cite{Buchmuller:1991tu,Ingelman:1993ve,Heusch:1993qu,Kersten:2007vk}
the Dirac mass matrix has rank~$1$,
\begin{equation} 
\label{eq:mDRank1}
	\mD = m \begin{pmatrix}
	        y_1 & y_2 & y_3 \\
	        \alpha y_1 & \alpha y_2 & \alpha y_3 \\
	        \beta y_1 & \beta y_2 & \beta y_3
	        \end{pmatrix} ,
\end{equation}
and if
\begin{equation} 
\label{eq:CancelCond}
	\frac{y_1^2}{M_1} + \frac{y_2^2}{M_2} + \frac{y_3^2}{M_3} = 0 \;,
\end{equation}
where $M_i$ are the singlet masses.
The mass parameters are defined in the basis where the singlet
mass matrix is diagonal.  The case of two RH neutrinos is analogous
\cite{Buchmuller:1990du,Datta:1991mf,Pilaftsis:1991ug},
while for four or more RH neutrinos there are additional possibilities.
The cancellation is valid to all orders in $\mD \mR^{-1}$.
The overall scale of the Yukawa couplings is not restricted by the
cancellation condition \eqref{eq:CancelCond} and hence allowed to be
large enough to make the detection of RH neutrinos possible.
The only relevant constraint is the experimental bound on the mixing
\begin{equation} \label{eq:nuNMixing}
	V = \mD \mR^{-1}
\end{equation}
between active and singlet neutrinos, \cite{Antusch:2006vw} 
\begin{equation}
	\sum_i |V_{\alpha i}|^2 \lesssim 0.01 \quad (\alpha = e, \mu, \tau) \;.
\end{equation}

\subsection{Underlying Symmetries}
\label{sec:Symmetries} 
Without a symmetry motivation, the cancellation condition
\eqref{eq:CancelCond} amounts to severe fine-tuning and is unstable
against radiative corrections.  Let us therefore discuss symmetries
leading to the cancellation.  We will restrict ourselves to the case of
three singlets.
A well-known possibility is imposing lepton number conservation
\cite{Wyler:1982dd,Bernabeu:1987gr,Tommasini:1995ii,Pilaftsis:2004xx,Pilaftsis:2005rv}.
The assignment $L(\nu_\text{L})=L(\nuR^1)=-L(\nuR^2)=1$, $L(\nuR^3)=0$
implies
\begin{equation} \label{eq:MassMatricesDP}
	\mR = \begin{pmatrix}
	       0 & M & 0 \\
	       M & 0 & 0 \\
	       0 & 0 & M_3
	       \end{pmatrix} , \;
	\mD = m \begin{pmatrix}
	         a & 0 & 0 \\
	         b & 0 & 0 \\
	         c & 0 & 0
	         \end{pmatrix} .
\end{equation}
Two singlets form a Dirac neutrino with mass $M$, while the third one
decouples.

An important question is whether lepton number conservation is also a
necessary condition for the cancellation of light neutrino masses, i.e.\
whether the cancellation can result from a symmetry that does not
contain $L$ conservation.
One can show that there is always a conserved lepton number, if the
cancellation occurs and if all three singlets have equal masses 
\cite{Kersten:2007vk}.
Let us therefore consider the case where the singlets involved in the
cancellation, say $\nuR^1$ and $\nuR^2$, have different masses and where
the condition \eqref{eq:CancelCond} is imposed by a symmetry at the
energy scale $M_2$.  Below this scale, the symmetry is broken.  The
neutrino masses change due to the renormalisation group running.  The
contributions from the two singlets to $m_\nu$ run differently between
$M_1$ and $M_2$ in the SM \cite{Antusch:2005gp}, so that the
cancellation is destroyed.  A rough estimate yields  
\begin{equation} \label{eq:MnuFromRunning}
	m_\nu(M_1) \sim 10^{-4}\GeV \, \ln\frac{M_2}{M_1}
\end{equation}
at $M_1$, which is unacceptable unless $\nuR^1$ and $\nuR^2$ are degenerate.
Of course, this problem persists if also the third singlet contributes
to the cancellation.

Thus, the cancellation of light neutrino masses can only be
realised without fine-tuning, if the RH neutrinos involved in the
cancellation have equal masses, which implies lepton number
conservation.  Therefore, any symmetry leading to vanishing neutrino
masses via this cancellation
has to contain the corresponding U(1)$_L$ as a subgroup or accidental
symmetry.

\subsection{Small Perturbations}
\label{sec:Perturbations}
Non-zero masses for the light neutrinos are obtained by introducing
small lepton-number-violating entries in the mass matrices
\eqref{eq:MassMatricesDP}.  In the most general case,
\begin{equation} \label{eq:GeneralPertDirac}
	\mR = \begin{pmatrix}
	       \epsilon_1 M & M & \epsilon_{13} M \\
	       M & \epsilon_2 M & \epsilon_{23} M \\
	       \epsilon_{13} M & \epsilon_{23} M & M_3
	       \end{pmatrix} , \;
	\mD = m \begin{pmatrix}
	         a & \delta_a & \epsilon_a \\
	         b & \delta_b & \epsilon_b \\
	         c & \delta_c & \epsilon_c
	     \end{pmatrix} .
\end{equation}
The smallness of the observed neutrino masses leads to the restriction
\begin{align}
	\epsilon_2 \, , \, \delta_{a,b,c} \lesssim 10^{-10}
\end{align}
for $\max(a,b,c) \sim 1$, $m/M \sim 0.1$, $M \sim 100\GeV$ (as required
by observability of RH neutrinos at LHC
\cite{Han:2006ip,delAguila:2006dx,Bray:2007ru,delAguila:2007em}),
provided that there are no special relations between the small
parameters causing additional cancellations.  The perturbations
$\epsilon_{23}$ and $\epsilon_{a,b,c}$ appear quadratically in $m_\nu$
and are correspondingly less severely constrained.  Finally,
$\epsilon_1$ and $\epsilon_{13}$ do not lead to neutrino masses at the
tree level at all but do contribute via loop diagrams \cite{Pilaftsis:1991ug}, so
that they are only slightly less constrained than the other parameters.

The most general mass matrices of \Eqref{eq:GeneralPertDirac} contain
many free parameters, so that there is no clear imprint of the
considered setup in the light neutrino mass matrix.
A more interesting phenomenology is possible in constrained cases,
some of which have been considered earlier in the context of
leptogenesis \cite{Raidal:2004vt,Pilaftsis:2005rv}.  For example, if all
small parameters are of the same order of magnitude,
\begin{equation} \label{eq:mnuLeadingDP}
	m_\nu \approx \frac{m^2}{M} \left[
	 \epsilon_2 \, v v^T - (v v_\delta^T + v_\delta v^T) \right] ,
\end{equation}
where we have abbreviated the first and second column of $\mD$ by $v$
and $v_\delta$, respectively.  The light neutrino masses are strongly
hierarchical, since $m_\nu$ has rank $2$ and hence one vanishing
eigenvalue.  The large Yukawa couplings $a,b,c$ are determined by the
light neutrino masses and mixing parameters, which leads to predictions for correlations
between the branching ratios of different lepton-flavour-violating decays in
supersymmetric see-saw models \cite{Raidal:2004vt}.  Likewise, the
amplitudes of LFV processes at colliders are correlated, as we will
discuss shortly.

\section{Signals at Colliders}
\label{sec:Colliders}
A striking signature of RH neutrinos at colliders would be
lepton-number-violating (LNV) processes with like-sign charged leptons
in the final state \cite{Keung:1983uu}.  However, we have argued that all symmetries
guaranteeing the required suppression of the light neutrino masses
lead to the conservation of lepton number, so that
the amplitudes of such processes vanish.  Any $L$ violation
is severely restricted by the smallness of neutrino masses and can
therefore not lead to sizable amplitudes.  Consequently, in the absence
of fine-tuning, LNV signals are expected be unobservable.

Another option are events with different leptons such as
$\mu^-\tau^+$ in the final state, since these have a relatively small SM
background as well.  
According to \cite{delAguila:2007em}, such signals are unlikely
to be observable at LHC, however.
In the considered scenarios, the mechanism leading to the cancellation
of neutrino masses causes the terms in the corresponding
amplitudes to add up constructively, leading to
\begin{equation} \label{eq:ALFV}
	A_{\alpha\beta} \propto
	\frac{m^2}{M^2} (a , b , c)_\alpha (a^* , b^* , c^*)_\beta
\end{equation}
for the mass matrices of \Eqref{eq:GeneralPertDirac}, where
$\alpha\neq\beta$ denote the flavours of the charged leptons.
If the cross sections are large
enough for a detection at colliders, flavour-violating decays of charged
leptons mediated by the RH neutrinos should be observable in upcoming
experiments as well, since their amplitudes depend on the same
combination of parameters.
In the constrained case that yields \Eqref{eq:mnuLeadingDP},
$a,b,c$ can be determined from the light neutrino mass parameters, as
mentioned above, so that the ratios $A_{e\mu}/A_{e\tau}$ and
$A_{e\mu}/A_{\mu\tau}$ are predicted.

At the ILC, the resonant
production of RH neutrinos is possible for $|V|_{ei}\gtrsim0.01$
\cite{delAguila:2005mf,delAguila:2006dx}.  By observing the branching
ratios for the subsequent decays into charged leptons, one could
then determine the mixings of the heavy neutrinos with the different
left-handed doublets directly.

\section{Summary and Discussion}
\label{sec:Concl}

We have discussed the prospects for testing the see-saw mechanism of
neutrino mass generation in collider experiments.  We have assumed the
existence of right-handed neutrinos with masses close to the
electroweak scale (but no other new particles or interactions).  The
couplings of these neutrinos to the SM particles can only be large
enough to make their observation at colliders possible, if different
contributions to the light neutrino masses nearly cancel.  This
cancellation is then the main reason for the smallness of the observed
neutrino masses, while the see-saw mechanism plays only a minor role.
Therefore, we have to conclude that a direct test of the see-saw mechanism at
the LHC or the ILC is not possible.

If one defines the leading-order mass matrices in such a way that they
correspond to exactly vanishing light neutrino masses, non-zero masses
appear as a result of small perturbations of this structure.  One may
then ask whether these perturbations could have consequences for signals
at colliders and thus allow for a test of the mechanism of neutrino mass
generation.  Unfortunately, the smallness of the light neutrino masses
immediately tells us that all perturbations are tiny and therefore
irrelevant for collider signatures.  Thus, the answer to this second
question is negative, too.  Collider experiments are only sensitive to
the leading-order mass matrices which do not lead to neutrino masses.

As a consequence, a connection between collider physics and neutrino masses
can only be established, if the perturbations are introduced in such a
way that the leading-order parameters are related to the light neutrino
masses and mixings.  In the most general case, this is not possible
because there are too many free parameters.  Then collider physics
decouples completely from the light neutrino masses and their generation.

However, the situation is better in constrained set\-ups where only some
of the perturbations are present or dominant.  In the cases we
discussed, a strong mass hierarchy is expected.  To the extent that the
leading-order Yukawa couplings are fixed by the measured neutrino masses
and mixings, correlations between the branching ratios of
lepton-flavour-violating processes can be obtained.  This applies both
to reactions at colliders and to LFV decays of charged leptons.
Finally, $e^+e^-$ colliders may be able to determine the mixings of RH
neutrinos with the different flavours directly.
Pursuing all
these experimental options provides a chance to test constrained setups of the kind
we have described.  Of course, even in this optimistic case it is
impossible to exclude the existence of additional, very heavy RH
neutrinos contributing to neutrino masses via the standard see-saw
mechanism.

Without an underlying symmetry, the described cancellation of the light
neutrino masses amounts to severe fine-tuning.  We have therefore
discussed symmetry motivations.  We have argued that every symmetry
realising the cancellation has to include lepton number conservation.
Otherwise, the cancellation is unstable against radiative corrections,
so that fine-tuning is still required.

Thus, both lepton number violation and light neutrino masses arise due
to small perturbations of the leading-order mass matrices, and their
magnitudes are related.  Therefore, we expect lepton-number-violating
signals at colliders to be unobservable in untuned scenarios.  The cross
sections for lepton-flavour-violating processes are not suppressed, so
that LHC experiments might be able to observe such reactions.  If this
is the case, lepton flavour violation should also be observable in
decays of charged leptons in the near future.

\section*{Acknowledgements}
\label{sec:Ack}

I'd like to thank Alexei Smirnov for the collaboration on
\cite{Kersten:2007vk}, on which this talk was based.
This work was supported in part by the European Commission under
the RTN contract MRTN-CT-2004-503369.

\bibliographystyle{NewArXiv}
\bibliography{../../Neutrinos}

\end{document}